# Surface nano-patterning through styrene adsorption on Si(100)

*A. Calzolari*[*,1], *A. Ruini*[1], *M.J. Caldas*[1,2], *and E. Molinari*[1]

[1] INFM-CNR National Center on nanoStructures and bioSystems at Surfaces (S3), and Dipartimento di Fisica, Università di Modena e Reggio Emilia, Via Campi 213/A, 41100 Modena, Italy

[2] Istituto de Fisica, Universidade de Sao Paulo, 05508-900 Sao Paulo, Brazil

* calzolari.arrigo@unimore.it

**ABSTRACT**

We present an ab initio study of the structural and electronic properties of styrene molecules adsorbed on the dimerized Si(100) surface at different coverages, ranging from the single-molecule to the full monolayer. The adsorption mechanism primarily involves the vinyl group via a [2+2] cycloaddition process that leads to the formation of covalent Si-C bonds and a local surface derelaxation, while it leaves the phenyl group almost unperturbed. The investigation of the functionalized surface as a function of the coverage (e.g. 0.5 – 1 ML) and of the substrate reconstruction reveals two major effects. The first results from Si dimer-vinyl interaction and concerns the controlled variation of the energy bandgap of the interface. The second is associated to phenyl-phenyl interactions, which gives rise to a regular pattern of electronic wires at surface, stemming from the π–π coupling. These findings suggest a rationale for tailoring the surface nano-patterning of the surface, in a controlled way.

**KEYWORDS** (Word Style "BG_Keywords"). If you are submitting your paper to a journal that requires keywords, provide significant keywords to aid the reader in literature retrieval.

**MANUSCRIPT TEXT**

Molecular electronics is extensively developed as a part of a long-term strategy to overcome the miniaturization problems in today's silicon-oxide based devices. After the seminal paper by Aviram and Ratner,[1] several works[2] have demonstrated the capability of molecular-based structures to perform electrical functions. These "concept devices" are based on hybrid metal-molecule-metal interfaces, where molecules are the active element of the device, while the metal pads constitute both the structural support for the molecule and the leads of the circuit. Most recent experiments are based on the thiol/Au chemistry, that exploits the ability of –SH group to anchor the molecule to the gold surface. Despite their basic relevance, these model systems present serious drawbacks for large-scale industrial applications, related both to the cost of materials and to the difficult control of the S-Au coupling.[3]

A parallel route is emerging, based on the organic functionalization of semiconductor surfaces.[4] The deposition of organic molecules on semiconductors allows one to obtain nanostructured materials, whose properties may be tuned in controlled way. While molecular bonding on metals is typically non-site-specific, the localized and directional character of semiconductor bonds at surfaces is able to impart an ordered arrangement to the adsorbed molecules, making the substrate an intrinsic template for the growth of the molecular layer. This opens to the possibility of exploiting these novel materials for nanotechnology (nanoelectronics, nonlinear optics, optoelectronics, etc.), and bioengineering applications (sensor activity, molecular recognition, etc).[5] In this letter, we show that the electronic properties of the hybrid interface can be engineered by tuning the dosage conditions for the adsorption.

The (2x1) reconstructed (100) and the (111) silicon surfaces seem particularly suited to couple to the carbon atoms of the organic molecules.[6] However, since the surface reactivity is essentially ruled by the presence of dangling bonds at surface, most experiments were done using the hydrogen terminated Si substrates.[7-10] Such surface passivation at the same time prevents spurious oxidation processes and induces a barrier for chemisorption, so that further dynamical mechanisms are required to promote the reaction. This was the case, e.g. of the self-directed growth of styrene rows on H:Si(100)-(2x1) surface.[11-12] On the other hand, the clean Si(100) surface exhibits a Si=Si dimer-like motif, whose electronic properties are in close analogy to those of the carbon-carbon double bond (C=C) of alkenes. Recent experiments have demonstrated that unsaturated hydrocarbons (e.g. ethylene,[13] benzene,[14] cyclopentene,[15-16] etc.) may easily bond to Si=Si dimers via cycloaddition reactions, typically used in organic chemistry.[17] Moreover, the direct measurement of the transport properties through single molecules (e.g. styrene, cyclopentene, TEMPO)[16,18] on Si(100) has recently shown the possibility of realizing operating semiconductor-based molecular devices.[19]

Deposition of styrene molecules on Si(100) is a representative model, since styrene is constituted of two building blocks, i.e. the phenyl ($-C_6H_5$) and the vinyl ($-CH-CH_2$) group, that are the key-components of most conjugated molecular structures and organic polymers. Experimental STM images[18,20] show indeed that styrene is preferentially located along Si dimer rows, and the analysis of infrared adsorption[20] and thermal desorption[21] spectra suggests that the chemisorption of styrene primarily involves the vinyl group. Although the adsorption of benzene molecules on Si(100) has been reported,[14] the anchorage of styrene through the phenyl group results to be highly unfavourable.

Despite the increasing interest about alkene adsorption on Si(100) surface, only few studies address its theoretical modelling: most of existing works[22-27] dealt with the adsorption geometries on finite clusters or the reaction pathways, but generally did not include the study of the electronic properties. In this letter, we present an ab initio investigation of the adsorption of styrene molecules on Si(100)-(2x1) surfaces. By means of a periodic solid-state approach, we studied the structural and electronic properties of the styrene/Si interface in the case of both the single-molecule and the monolayer adsorption. Our results indicate that under appropriate adsorption conditions it is possible to control the conduction properties of the organic-molecule/inorganic-surface interface.

We performed state-of-the-art electronic structure calculations[28] based on Density Functional Theory (DFT), with the PW91 generalized gradient approximation for the exchange-correlation functional.[29] Electron-ion interactions were described using ultrasoft pseudopotentials,[30] and the single particle electronic wavefunctions were expanded in plane waves with a kinetic energy cutoff of 20Ry. The Si(100) surfaces were modeled by means of repeated slab supercells containing 6 atomic layers and ~16Å of vacuum. We simulated the (2x1) reconstruction of Si(100) in cells with different lateral periodicity, depending on the coverage: a large c(16x16) cell was used in the case of single molecule adsorption to isolate molecule in neighbor cells; and the p(2x2) cell was used in the case of monolayer configurations (see Fig.1). Styrene was adsorbed onto one surface of the slab, while a monolayer of H atoms was used to saturate the dangling bonds (DBs) on the back side. All structures were relaxed until forces on all atoms were lower than 0.03 eV/ Å.

Our results for the clean surface reproduce well the existing experimental[6] and theoretical[31] results: the outermost atoms assemble in buckled dimers along the [011] direction (see Fig.1). The dimerization process corresponds to the formation of a full σ and a weak π double bond. The Si=Si tilting imparts a "zwitterionic" character to the dimer, associated to a charge transfer from the "down" to the "up" atom. Accordingly, the lowest unoccupied molecular orbital (LUMO) and the highest occupied molecular orbital (HOMO) are surface states, localized around the "down" and the "up" atom respectively.

The study of single molecule adsorption provides a great deal of insight into the mechanisms that rule the bonding properties at the interface. On the basis of the experimental evidence,[18,20,21] the starting configuration for the ab initio relaxation was obtained by orienting the vinyl group of styrene in proximity to a Si=Si dimer (marked in dark red in Fig.1). The adsorbed styrene results bridge-bonded to the Si dimer, with the vinyl group only slightly misaligned with respect to the dimer below. The surface does not exhibit structural distortions, except for the adsorption site where the substrate strongly derelaxes, removing the buckling only of the dimer involved in the bonding process (Fig. 1b). The final product is formally equivalent to a [2+2] cycloaddition reaction,[17] in which the π orbitals of both the alkene C=C and the Si=Si dimer couple and create two covalent Si-C σ bonds in a four membered Si-Si-C-C ring (Fig. 1c). According to the Woodward-Hoffman selection rules,[4,17] [2+2] cycloaddition reactions between truly double bonded members (e.g. alkenes) should be symmetry forbidden. However, the solid-state effects, responsible for the dimer buckling, break the orbital symmetry of the Si double bond allowing the reaction to occur. The formation of Si-C bonds implies a uniform charge redistribution, which removes the tilting relaxation of the clean dimer (Fig. 1b). Since the binding mechanism involves the vinyl group and a single dimer, the counter-relaxation occurs locally at the adsorption site only, leaving the rest of the surface as well as the phenyl group almost unperturbed.

The electronic structure of the styrene/Si(100) interface confirms the high selectivity of the adsorption process. In Figure 2a we compare the total density of states (DOS) of the Si surface with (thin black line) and without (dotted orange line) the presence of the single styrene molecule. The projection on the molecular states (red area) primarily affects the low energy range of the spectrum, adding new peaks to the original Si-DOS (not shown). The region near the Fermi energy is dominated by the Si states: the HOMO and the LUMO peaks maintain the features of the corresponding states in the clean surface, i.e. localized on the "up" (HOMO) and "down" (LUMO) atoms respectively. The unique exception is the bonding dimer, where a node of the HOMO and LUMO wavefunctions is observed (supplementary 1a,b). To find out the electronic states involving the adsorption site we have to move away from the Fermi energy towards the continuum of states of the Si surface. The thick black line in Fig. 2 represents the DOS projection onto the C atoms of the former vinyl group of styrene and the underlying Si dimer. The peaks closest to the Fermi level are located at −1.0 eV (labeled [2+2] in Fig. 2) and +1.8 eV (labeled [2+2]*), and correspond to the bonding and antibonding states resulting from the cycloaddition reaction (supplementary 1c,d). The formation of the Si-C bonds –energetically more stable than the Si=Si double bond– is responsible for the downward shift of the [2+2] peak. Since the reaction involves only a single dimer (1 out of 16 in our simulation), the adsorption of a single molecule does not significantly modify the electronic properties of the surface around the Fermi energy.

The electronic structure at the edge of the valence band is significantly represented in the simulated STM image (supplementary 2a), that was obtained within the Tersoff-Hamman approximation[32] at −2.0 eV bias. We distinguish the uniform pattern of spots related to the upper part of the Si dimers and the bright protrusion centred on the styrene and on the dimer beneath. It is worth noting that the phenyl group, which does not participate in the adsorption process, maintains its original aromatic character.

The simulated STM image agrees well with the experimental results.[18,20] However, at low dosage conditions it has been observed[20] that molecules do not always have the same orientation with respect to the dimer rows, the phenyl group being found on both the right and left sides of a row. To justify this finding, we considered an alternative reconstruction of the Si(100) surface. It is known that Si(100) may undergo different reconstructions depending on the experimental conditions;[31] here we simulated the Si(100)-(2x2) surface, which is characterized by parallel rows of alternating buckled dimers along the [011] direction. Results for the adsorption of a single styrene onto this surface are displayed in supplementary 2b. The alternating spots in the STM image reflect the alternating orientation of the Si=Si dimers, and no significant changes are observed close to the bonded molecule. The adsorption mechanism is the same as for the (2x1) case: since the [2+2] cycloaddition involves a single Si=Si site, the relative orientation of the dimers does not modify the bonding properties at the interface, driving only the lateral orientation of the phenyl group.

Increasing the dosage, styrene adsorbs on Si(100)-(2x1) at a saturation coverage of 1 monolayer (ML), which corresponds to one styrene for every surface dimer.[21] The self-assembled overlayer is highly ordered and oriented along the dimer rows, with a intermolecular distance of 3.8 Å along the [011] direction, induced by the substrate periodicity. Experimental observations[3,4] show that spatial arrangement of self-assembled monolayers (SAMs) may be strongly affected by non-bonding interactions such as Van der Waals (VdW) forces. Since standard DFT does not include these effects, we used the well established polymer consistent force field (PCFF)[33] to reach the equilibrium structures. The starting atomic configurations at 1ML coverage were obtained saturating each Si=Si dimer with a styrene molecule. We first optimized the structure at the PCFF level (which treats electrostatic and VdW interactions), and then calculated the corresponding electronic structure at the DFT level, keeping the atoms fixed at the relaxed geometry.

We considered the styrene adsorption on the Si(100)-(2x1) and the -(2x2) surfaces. We labeled the former system $[1ML@(2x1)]_{(2x2)}$ and the latter $[1ML@(2x2)]_{(2x2)}$, where the internal parenthesis refers to the original substrate reconstruction and the subscript is the periodicity of the supercell.

In both cases the molecule/surface bonding is referable to a [2+2] cycloaddition reaction, through the formation of a four membered ring for each adsorbed molecule. However, passing from the single molecule to the monolayer configuration, the different Si(100) reconstructions may result in different steric couplings among the molecules. The characteristics of the dimerized surface impart a specific order to the overlayer, acting as a programmable template for the growth of the organic material. Thus the monotonic buckling of (2x1) reconstruction leads to a parallel arrangement of the phenyl groups, that align vertically along one side of the dimer row (Fig. 3a). The (2x2) reconstruction causes, instead, a zig-zag alternation of the aromatic rings, that arrange in a herring-bone structure along the [011] direction (Fig. 3b).

Despite the different spatial arrangement of the overlayers, the two systems present interesting similarities in their electronic sturctures. In particular, the analysis of the functionalized surfaces reveals two major effects, related to Si dimer-vinyl and to phenyl-phenyl interactions respectively. The former concerns the bonding properties at the Si/molecule interface. Since the styrene chemisorption is site-specific and highly localized, the monolayer configuration results to be a simple superposition of single adsorption events. Following the lines described above, for each bonded molecule we observe the breaking of Si-dimer double bonds and the formation of bonding ([2+2] in Fig.2) and anti-bonding ([2+2]*) Si-C orbitals, lying in the region of continuum bulk states of the surface. In the limit of full coverage, where all dimers are involved, we observe the complete suppression of the original HOMO and LUMO peaks of the clean surface, which were representative of the Si=Si double bonds. The HOMO and LUMO peaks of the fully covered surface are the result of the saturation of the Si-surface dangling bonds through the hybridisation between Si-bulk and Si-C [2+2] states, as shown in Figure 2b and supplemetary 3 for the [1ML@(2x1)]$_{(2x2)}$ structure. Therefore, the overall effect of the overlayer formation is the opening of the bandgap.

To understand the origin of this bandgap opening, we compare (Fig.4) the DOS of clean -(2x1) (solid line) and -(2x2) (dashed line) reconstructions (panel a) with the corresponding functionalized surfaces (panel b). The two curves for the clean surfaces (panel a) are quite similar (the slight discrepancies are due to the specific relaxation mechanisms): they are both characterized by a small gap caused by the presence of surface states in the original Si gap.[34] In the monolayer configuration (panel b) these states completely disappear, changing the bandgap. It is worth noting that the DOS's of panel (b) are almost identical, indicating that the bandgap variation depends only on the Si-dimer saturation (i.e. the Si-vinyl interaction) and not on the details of the starting surface reconstruction or of the overlayer arrangement. To support this hypothesis we studied the case of ethylene ($C_2H_4$) and mono hydride adsorption on the Si(100)-(2x1) surface at 1ML coverage. Figure 4c shows the same gap enlargement as in the styrene case; the further features related to the specific adsorbate (e.g. the vinyl group in ethylene) affect the DOS only in other energy regions. This confirms that gap opening is ruled only by the saturation process of the exposed dimers, and not by other fragments (e.g. the phenyl group of styrene).

Along these lines, we considered further intermediate configurations, where only a part of the original Si surface states are saturated. We studied in particular two reconstructions of styrene adsorbed on Si(100)-(2x1) at 0.5 ML coverage. The first (labeled [0.5ML@(2x1)]$_{(2x2)}$) is a (2x2) structure, where each dimer row is constituted of the alternation of buckled dimers and styrene molecules along the [011] direction (supplementary 4a). The second (labeled [0.5ML@(2x1)]$_{(4x1)}$ ) has a (4x1) periodicity, and consists in the alternation along the [01-1] direction of an unperturbed Si-dimer row and a styrene-saturated row (supplementary 4b). In both cases 50% of Si-dimers persist in the clean surface configuration. The resulting DOS's are shown in Fig. 4 (bottom panel). We focused on the modification of the frontier orbitals with respect to the Si-dimer surface states of the clean substrate (Fig. 4e). The vertical lines indicate the large gap at saturation conditions. The [0.5ML@(2x1)]$_{(4x1)}$ structure presents common features to both the clean and the fully saturated system: the energy gap is similar to that of the clean surface. The HOMO and LUMO peaks (vertical arrows) correspond to the unperturbed Si-dimers and have a halved intensity. The presence of the styrene introduces spectral features about −1eV, outside the full saturation gap (vertical lines), as in the 1ML case. On the contrary, the on-row alternation in the [0.5ML@(2x1)]$_{(2x2)}$ system breaks the intrinsic coupling among dimers, suppressing the HOMO peak but leaving states in the region of the LUMO peak of the clean surface. Indeed, the resulting bandgap is larger than for the clean surface but smaller than for the fully covered system. It is worth noting that the HOMO peak is more sensitive to the adsorption of the molecule; on the contrary, the LUMO maintains its original character in the two reconstructions (Fig.4e,f). In both cases (see supplementary 4) the LUMO states are localized around the unsaturated dimers, and the LUMO+1 are bulk-like silicon states, delocalised on the whole structure, but the molecule. On the basis of these results, we suggest that the functionalised surfaces at submonolayer coverage –even with less ordered overlayers- may act as electron acceptors, and promote the charge transport through the silicon conduction states.

We conclude that it should be possible to tune the bandgap and the conduction properties of the interface by controlling the coverage of the adsorbed molecules.

However, the limitation of the adsorption process only to the passivation of Si-dimers is not sufficient to account for the differences in the functionalization of the surface, with respect to different adsorbates (styrene, ethylene, hydrogen, etc). The observed differences are instead related to the presence of other functional groups (phenyl in the present case), that may drive both the spatial arrangement of the molecules at surface and the formation of new electronic features. We analysed the effects of the phenyl-phenyl interactions. Passing from the single molecule to the monolayer configuration we observe the broadening of the peaks deriving from the aromatic ring. This is a signature of the $\pi-\pi$ interaction between the assembled molecules that tends to create delocalized orbitals at surface, as shown in Supplementary 2 in the case of $[1ML@(2x1)]_{(2x2)}$ structure. The closer is the packing, the higher is the superposition of the molecular orbitals. The formation of dispersive states may in principle enhance the in-plane transport properties of the organic overlayer, provided the bandgap variations described above and a proper doping.

The specific relative orientation among the aromatic rings plays a crucial role in the expected efficiency for intermolecular transport. Despite their similar bandstructure properties (see Fig. 5(a)), the 1 ML coverages on the (2x1) and (2x2) reconstructions give rise to cofacial and herringbone arrangements respectively, which in principle turn out different in-plane transport efficiency: the direct $\pi-\pi$ organization typically results in electronic transfer integrals that are one order of magnitude larger than in the herringbone structure.[35] Besides the 2x1 reconstruction for the 1 ML coverage, we therefore argue that also the 4x1 reconstructions for the 0.5 ML coverage might originate satisfactory hole transport performances along the styrene-saturated 1D rows.

The different arrangements of the overlayer may be distinguished through the analysis of the delocalized π-channels, by means of STM experiments. In Figs. 3c-f we compare the calculated STM images for the target systems: the current signal is stronger in the [1ML@(2x1)]$_{(2x2)}$ structure, which realizes the maximum π−π superposition. The π-channels expand parallel to the dimer rows, while they are almost localized in perpendicular direction (see Figs. 4c,f). This implies that the directional Si-dimer motif of the clean surface drives also the formation of an ordered pattern of electronic wires on top of the surface, that may be appealing for further nanoscale applications. It is worth noting that these electronic features, being induced by the phenyl-phenyl interaction, can not be obtained with other adsorbates such as hydrogen or ethylene.

In conclusion, we studied the effects of styrene adsorption on dimerized Si(100) surfaces. We described the bonding mechanism in terms of the electronic structure of the system, highlighting the effects of the Si-C bond formation on the overall properties at different submonolayer coverage configurations. The localized and directional nature of the Si-dimer rows governs both the atomic structure and the conduction properties of the overlayer. Our results show different regimes as a function of substrate reconstruction and of the dosage separately: the former is responsible for the final arrangement of the molecular monolayer (e.g. cofacial vs. herringbone); the latter drives the conduction properties and the electron distribution of the surfaces. Indeed, at high coverage (e.g. [1ML@(2x1)]$_{(2x2)}$) the effective π−π coupling may induce hole transport along the molecular layer, while at submonolyer regimes the electron transport through the Si states is favored. Thus, small organic molecules (such as styrene) seem to be attractive candidates to modify both the structural and the electronic properties of semiconductor surfaces, paving the way for novel applications in the field of semiconductor-based molecular electronics.

This work was supported in part by: MIUR (Italy) through grant FIRB-Nomade and by INFM through "Progetto calcolo parallelo". We thank Clotilde Cucinotta, and Rosa Di Felice for fruitful discussions.

[34] We are well aware of the problems related to the evaluation of the energy bandgap in supercell DFT calculations; namely two antithetic effects: the underestimation of bandgap typical of the DFT approach, and the gap opening due to the finite-size of the surface slab. On the other hand, we are interested in the mechanisms that rule the gap control at interface, while the exact evaluation of the bandgap for comparison with the experiments goes beyond the aim of this work.

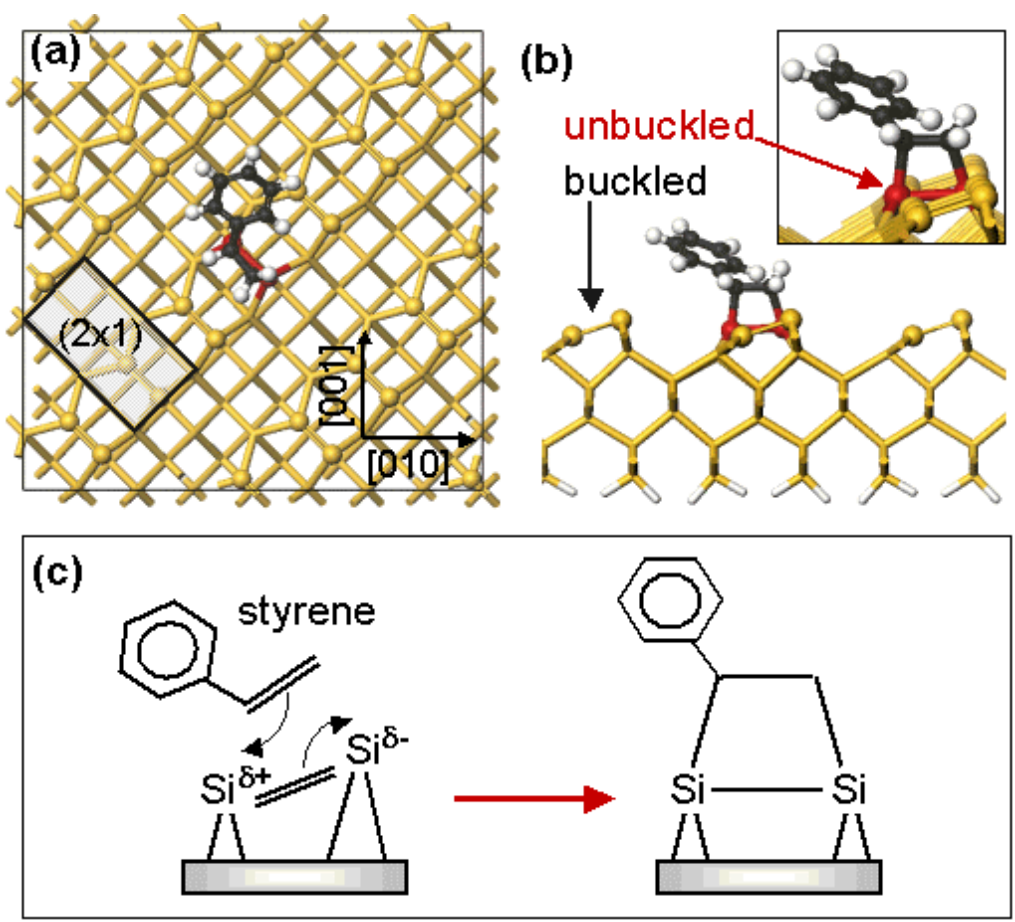

**Figure 1.** Top (a) and side-view (b) of a single styrene molecule adsorbed on a Si(110)-(2x1) surface, simulated in a c(16x16) cell (shaded area defines the primitive (2x1) cell). Balls identify the atoms of the first layer. Si-dimer rows are aligned along the [011] direction, adsorption site is highlighted in red. The inset zooms into the locally unbuckled adsorption site. (c) Schematic illustration of [2+2] cycloaddition of styrene on Si(100)-2x1 surface. The reaction, formally forbidden by symmetry constraints, is instead facilitated by the asymmetry of the surface reconstruction, with the consequent charge transfer ($\delta$) between Si atoms.

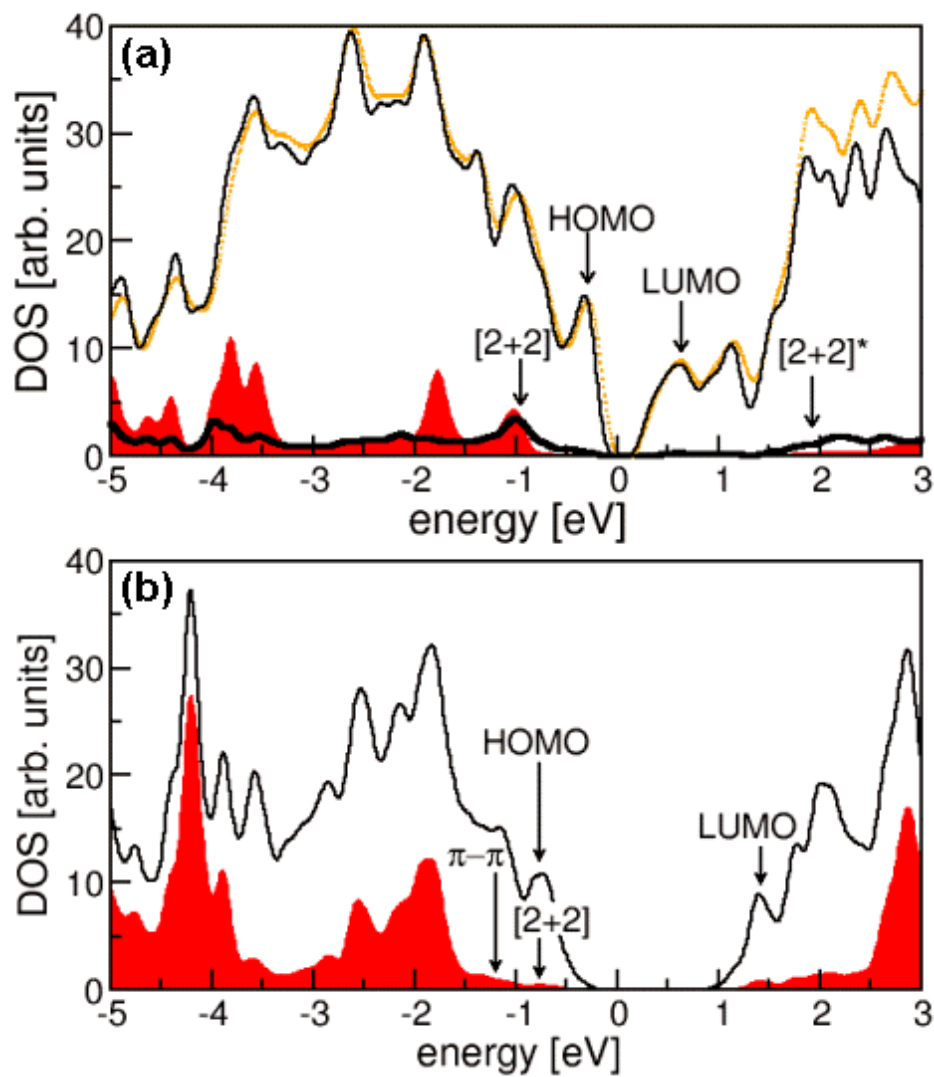

**Figure 2.** (a) Single molecule adsorption. Total DOS (thin black) and projection on styrene atoms (red area) from Styrene/Si(100)-(2x1) interface. Dotted orange line is the total DOS of clean surface, thick line is the projection on atoms involved in cycloaddition reaction. Total DOS for clean and adsorbed surfaces are scaled by a factor 0.45 in order to magnify the component projections (red area and thick line). (b) [1ML@(2x1)]$_{(2x2)}$ configuration. Total DOS (black line) and projection on styrene atoms (red area) from Styrene/Si(100)-(2x1) interface at 1ML coverage. All curves were aligned to the bottom of silicon valence band. Zero energy reference corresponds to the Fermi level of the (2x1) clean surface.

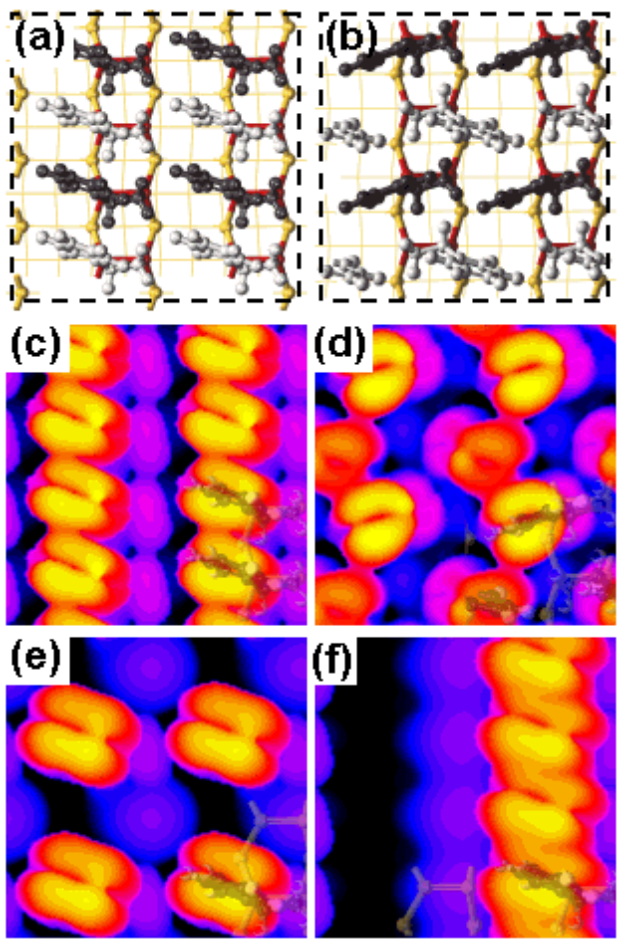

**Figure 3.** Top panel: optimized atomic geometries for styrene molecules adsorbed on Si(100) -(2x1) (a) and -(2x2) (b) reconstructions at 1ML coverage, namely [1ML@(2x1)]$_{(2x2)}$ and [1ML@(2x1)]$_{(2x2)}$ configurations respectively. Ball-stick rendering corresponds to styrene molecules and to the two outermost Si layers, thin lines to planes beneath. Red atoms mark the unbuckled Si dimers. Styrene molecules are alternately colored (light vs dark gray) for clarity. Middle Panel: calculated STM images for 1ML coverage on (2x1) (c) and (2x2) (d) surface reconstructions, corresponding to the structures (a) and (b) respectively. Bottom panel: 0.5 ML coverage configurations on Si(100)-(2x1) substrate with (2x2) (e) and (4x1) (f) lateral periodicity. Atoms of unit cells are superimposed for clarity.

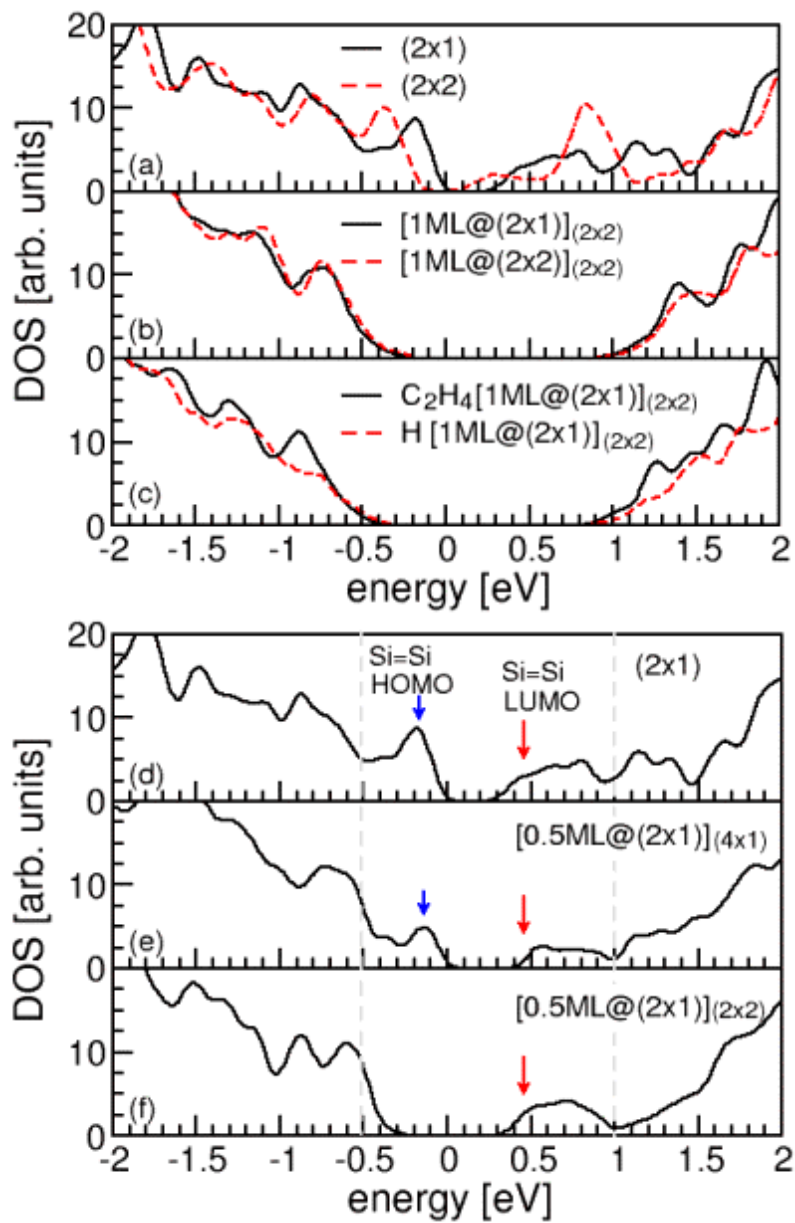

**Figure 4.** Top panel: density of states near the band gaps for (a) clean Si(100)-(2x1) and (2x2) reconstructions; (b) styrene adsorption on (2x1) and (2x2) Si substrate at 1ML coverage; panel (c) ethylene (C$_2$H$_4$) molecule and H atom adsorption on Si(100)-(2x1) at full monolayer coverage. Bottom panel: modification of Si-dimer HOMO (blue arrows) and LUMO (red arrows) peaks of the clean (2x1) surface (d) at 0.5 ML coverage, for the (4x1) (e) and (2x2) (f) reconstructions. Vertical dashed lines correspond to the energy gap at saturation (see top panel). All curves were aligned to the bottom of silicon valence band; zero energy reference corresponds to the Fermi level of the (2x1) clean surface.

# Surface nano-patterning through styrene adsorption on Si(100)

*A. Calzolari* [*,1], *A. Ruini* [1], *M.J. Caldas* [1,2], *and E. Molinari* [1]

[1] INFM-CNR National Center on nanoStructures and bioSystems at Surfaces (S3), and Dipartimento di Fisica, Università di Modena e Reggio Emilia, Via Campi 213/A, 41100 Modena, Italy

[2] Istituto de Fisica, Universidade de Sao Paulo, 05508-900 Sao Paulo, Brazil

## Supplementary material

*CORRESPONDING AUTHOR FOOTNOTE INFM-CNR National Center for nanoStructures and bioSystems at Surfaces (S3), c/o Dipartimento di Fisica, Università di Modena e Reggio Emilia, Via Campi 213/A, 41100 Modena, Italy. Phone: +39-059-2055627. Fax: +39-059-2055651. e-mail: calzolari.arrigo@unimore.it

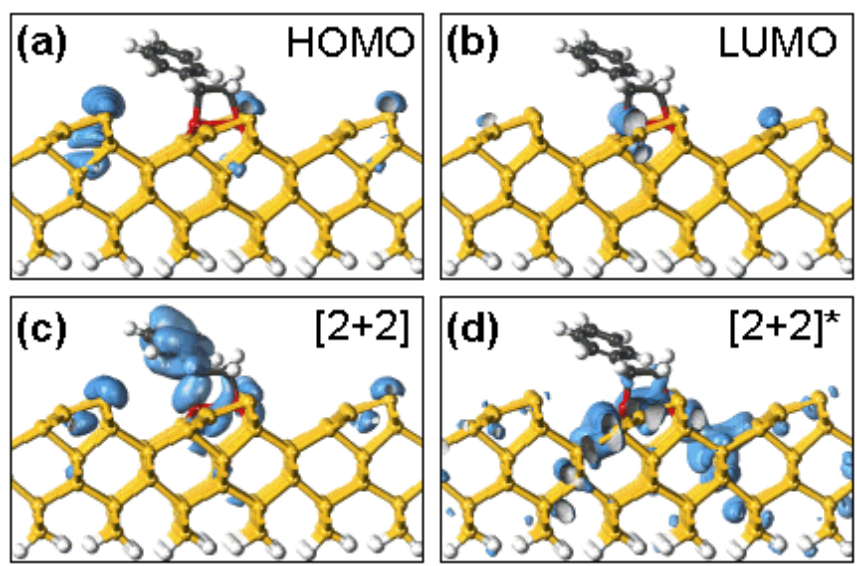

**Supplementary1** Single molecule adsorption. Isosurface plots of selected single-particle states (side view). Labels refer to Figure 2a.

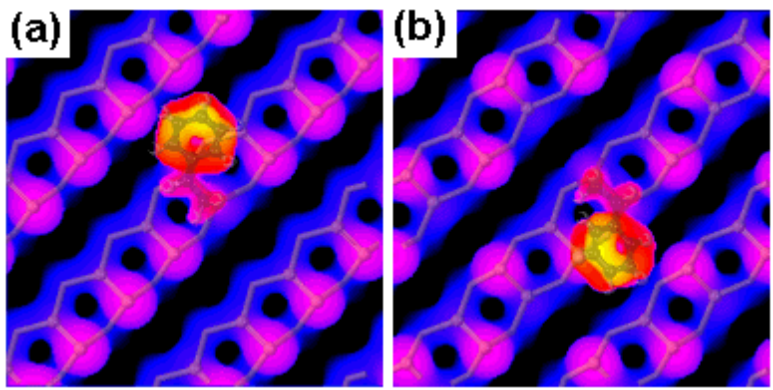

**Supplementary 2.** Calculated STM images for single styrene adsorption on Si(100) (a) (2x1) and (b) (2x2) surfaces. Atomic positions of the styrene molecule and of the two outermost Si layers are superimposed for clarity.

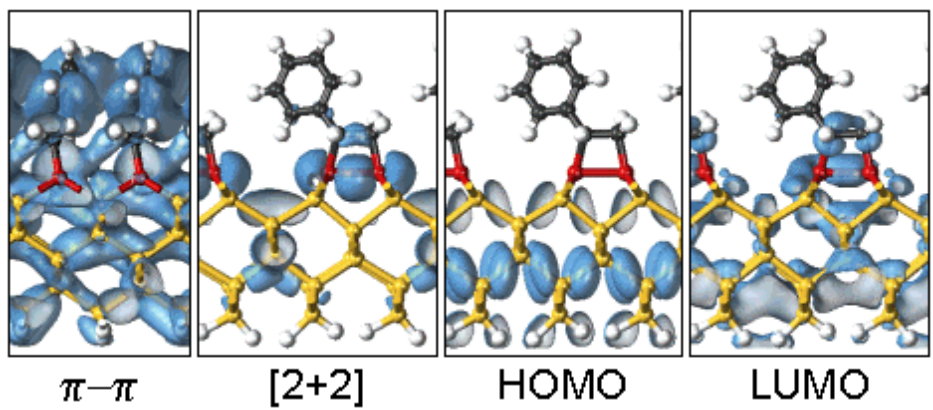

**Supplementary 3** [1ML@(2x1)]$_{(2x2)}$ configuration. Isosurface plots of selected single-particle states (side views). Labels refer to Figure 2b.

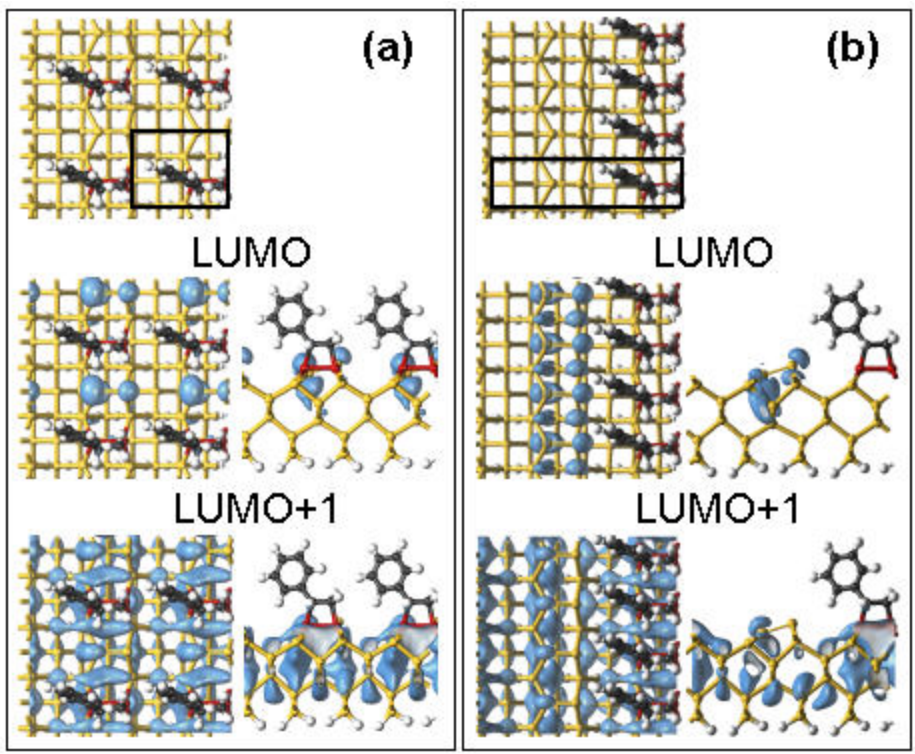

**Supplementary 4** Styrene adsorption on Si(100)-(2x1) surface at 0.5 ML. (a) [0.5ML@(2x1)]$_{(2x2)}$ and (b) [0.5ML@(2x1)]$_{(4x1)}$ configuration. Top view (left) and side view (right) of isosurface plots of LUMO (middle panel) and LUMO+1 (bottom panel). Dark lines label the unitary (2x2) (left panel) and (4x1) (right panel) cell used in the simulations. Unitary cells have been replicated for clarity.